\documentclass[10pt,letterpaper]{article}
\usepackage[top=0.85in,left=2.75in,footskip=0.75in,marginparwidth=2in]{geometry}

\usepackage{cite}

\usepackage{textcomp,marvosym}

\usepackage{amsmath}
\usepackage{textgreek}

\usepackage{nameref,hyperref}

\usepackage[right]{lineno}

\usepackage{microtype}
\DisableLigatures[f]{encoding = *, family = * }

\raggedright
\usepackage{changepage}

\usepackage[aboveskip=1pt,labelfont=bf,labelsep=period,singlelinecheck=off]{caption}

\makeatletter
\renewcommand{\@biblabel}[1]{\quad#1.}
\makeatother

\usepackage{lastpage,fancyhdr,graphicx}
\usepackage{epstopdf}
\fancyheadoffset[L]{2.25in}
\fancyfootoffset[L]{2.25in}

\usepackage{color}

\definecolor{Gray}{gray}{.25}

\usepackage{graphicx}

\usepackage{sidecap}

\usepackage{wrapfig}
\usepackage[pscoord]{eso-pic}
\usepackage[fulladjust]{marginnote}
\reversemarginpar

\begin{document}
\vspace*{0.35in}

\begin{flushleft}
{\Large
\textbf\newline{Impacts of Antarctic fast dynamics on sea-level projections and coastal flood defense}
}
\newline
\\
Tony E. Wong\textsuperscript{1,*},
Alexander M.R. Bakker\textsuperscript{1,\textdagger},
and Klaus Keller\textsuperscript{1,2,3}
\\
\bigskip
\bf{1} Earth and Environmental Systems Institute, Pennsylvania State University, University Park, PA 16802, USA
\\
\bf{2} Department of Geosciences, Pennsylvania State University, University Park, PA 16802, USA
\\
\bf{3} Department of Engineering and Public Policy, Carnegie Mellon University, Pittsburgh, PA 15289, USA
\\
\bf{\textdagger} Now at: Rijkswaterstaat, Ministry of Infrastructure and Environment, Utrecht, Netherlands
\\
\bigskip
* Corresponding author: Tony E. Wong, 2217 EESB Pennsylvania State University, University Park, PA, 16802, USA; twong@psu.edu

\end{flushleft}

\section*{Abstract}
Strategies to manage the risks posed by future sea-level rise hinge on a sound characterization of the inherent uncertainties. One of the major uncertainties is the possible rapid disintegration of large fractions of the Antarctic ice sheet in response to rising global temperatures. This could potentially lead to several meters of sea-level rise during the next few centuries. Previous studies have typically been silent on two coupled questions: (i) What are probabilistic estimates of this ``fast dynamics'' contribution to sea-level rise? (ii) What are the implications for strategies to manage coastal flooding risks? Here, we present probabilistic hindcasts and projections of sea-level rise to 2100. The fast dynamics mechanism is approximated by a simple parameterization, designed to allow for a careful quantification of the uncertainty in its contribution to sea-level rise. We estimate that global temperature increases ranging from 1.9 to 3.1~\textdegree C coincide with fast Antarctic disintegration, and these contributions account for sea-level rise of 21-74 centimeters this century (5-95\% range, Representative Concentration Pathway 8.5). We use a simple cost-benefit analysis of coastal defense to demonstrate in a didactic exercise how neglecting this mechanism and associated uncertainty can (i) lead to strategies which fall sizably short of protection targets and (ii) increase the expected net costs.

\section{Introduction}
\paragraph{}
Rising sea levels drive severe risks for many coastal communities~\cite{nicholls2010,hinkel2015}, The design of coastal defense strategies can hinge critically on future sea-level projections. Deriving probabilistic projections of sea-level rise poses nontrivial challenges, as they must account for a complex mixture of uncertainties surrounding the models and data employed~\cite{bakker2017a,kopp2016,mengel2016}. One important source of uncertainty is driven by the potential disintegration of the West Antarctic ice sheet (WAIS) and general Antarctic ice sheet (AIS) fast dynamics~\cite{alley2005,bamber2013b}. Potentially important mechanisms that contribute to these AIS fast dynamics include marine ice sheet instability (MISI), hydrofracturing and ice cliff instability~\cite{pollard2015,deconto2016}. To better understand the local coastal defense decisions that must be made, sea level projections must include all major contributions to local sea level~\cite{cazenave2014}. Additionally, geological factors affecting local relative sea level changes, such as sediment compaction, ground water, oil and gas extraction, and glacial isostatic adjustment – all of which contribute to potential land subsidence~\cite{Jones2016} – must be taken into account.

\paragraph{}
Recent studies have made considerable progress towards understanding these mechanisms, including through process-based modeling~\cite{joughin2014,pollard2015,deconto2016}, probabilistic projections and statistical modeling~\cite{little2013b,diaz2016wais}, and expert assessment~\cite{bamber2013b}. For example, recent work successfully constrained the AIS/WAIS dynamics by synthesizing expert assessment with probabilistic inversion and projections~\cite{Oppenheimer2016}. Here, we take an alternative probabilistic modeling approach. The current nexus of paleoclimatic as well as modern observations, more complete models and the ability to fuse models and data has presented the opportunity to produce probabilistic sea-level rise projections that include the effects of the AIS fast dynamics, constrained using paleoclimatic as well as modern observational data.

\paragraph{}
Previous probabilistic projections of sea-level rise (e.g., ~\cite{Kopp2014,Jackson2016,kopp2016,mengel2016}) have typically excluded a calibrated parameterization for the potential fast Antarctic ice sheet contributions to sea level. Jevrejeva et al.~\cite{Jevrejeva2014} combined process-based modeling with expert assessment for the fast Antarctic dynamics~\cite{bamber2013b} to find an upper limit of sea-level rise this century of 180 cm. The need for probabilistic projections that account for the Antarctic fast dynamical sea level contributions has been largely unfulfilled, but not unnoticed (see the discussion in ~\cite{Oppenheimer2016}). Here, we implement a simple, mechanistically-motivated parameterization for the AIS fast dynamics contribution to sea-level rise. Other studies have incorporated these fast dynamics effects into projections of sea-level rise based on statistical modeling and emulation of more detailed process-based models~\cite{deconto2016,Kopp2017,LeBars2017}. The incorporation of the fast Antarctic dynamics into probabilistic projections of sea-level rise, specifically through a simple physically-motivated model calibrated directly to paleoclimatic data, is the key advance of the present study.

\paragraph{}
Our projections of sea-level rise for this century are higher than previous projections, so the risks of coastal flooding are also likely higher. The sea-level projections are used in an intentionally simple and illustrative cost-benefit analysis method to quantify the impacts of the new scientific findings on coastal flood risk and strategies to manage these risks~\cite{dantzig1956}. Specifically, we evaluate the flood protection strategy for the north-central levee ring in New Orleans, Louisiana~\cite{jonkman2009b}, assuming policy-makers either use or neglect the additional AIS fast dynamics contributions to future sea-level rise. We conclude with a comparison of the two strategies, revealing the impacts of neglecting the fast dynamics. We stress that these coastal defense results should be viewed as a didactic exercise, demonstrating one sensitivity of flood protection strategies and costs to an improved representation of the Antarctic fast dynamics.

\section{Methods}
\subsection{Sea-Level Rise}
\paragraph{}
We employ and expand upon a model framework that has been previously applied for probabilistic projections of sea-level rise~\cite{bakker2017a}. This model has recently been made available as the Building blocks for Relevant Ice and Climate Knowledge (BRICK) model v0.1 to simulate global mean surface temperature, ocean heat uptake, global mean sea level and its contributions from the Antarctic ice sheet, Greenland ice sheet, thermal expansion and glaciers and small ice caps~\cite{Wong2017}. BRICK uses a semi-empirical modeling approach, combining a platform of previously published models. The model is described in greater detail by Wong et al.~\cite{Wong2017}, so we only provide an overview here.

\paragraph{}
Global mean surface temperature and ocean heat uptake are simulated by the zero-dimensional Diffusion-Ocean-Energy balance CLIMate model DOECLIM~\cite{kriegler2005}. DOECLIM is a zero-dimensional energy balance model coupled to a three-layer, one-dimensional diffusive ocean model. The input required to force DOECLIM is the radiative forcing time series (W m$^{-2}$), which is provided as in previous studies using DOECLIM~\cite{urban2010,Urban2014}. We use a one-year time step, and the output global mean surface temperature couples to the sea level sub-models representing individual major sea level contributions. All sea level is presented relative to 1986-2005 mean.

The Greenland ice sheet is represented by the Simple Ice-sheet Model for Projecting Large Ensembles, or SIMPLE~\cite{bakker2016a}. SIMPLE first estimates an equilibrium Greenland ice sheet volume ($V_{eq,GIS}$), given an anomaly in global mean temperature ($T_g$), as well as the e-folding time-scale of the ice sheet volume as it exponentially relaxes towards this equilibrium volume ($\tau_{GIS}$).
\begin{center}
\begin{equation}
\label{eq:veq_gis}
V_{eq,GIS}(t) = c_{GIS} T_g(t) + b_{GIS}
\end{equation}
\end{center}
\begin{center}
\begin{equation}
\label{eq:tau_gis}
\frac{1}{\tau_{GIS}(t)} = \alpha_{GIS} T_g(t) + \beta_{GIS}
\end{equation}
\end{center}
In Equations \ref{eq:veq_gis} and \ref{eq:tau_gis}, $t$ represents time (in years) $c_{GIS}$ is the equilibrium ice sheet volume sensitivity to temperature (m SLE~\textdegree C$^{-1}$), $b_{GIS}$ is the equilibrium ice sheet volume for zero temperature anomaly (m SLE), $\alpha_{GIS}$ is the temperature sensitivity of the e-folding ice sheet response time-scale (~\textdegree C$^{-1}$ y$^{-1}$), and $\beta_{GIS}$ is the equilibrium response time-scale (y$^{-1}$). These quantities are uncertain model parameters, which we estimate as described in Wong et al. \cite{Wong2017} and briefly in Section 2.3. The change in Greenland ice sheet volume ($V_{GIS}$) can then be written as
\begin{center}
\begin{equation}
\label{eq:v_gis}
\frac{dV_{GIS}}{dt}(t) = \frac{1}{\tau_{GIS}(t)} (V_{eq,GIS}(t) - V_{GIS}(t)).
\end{equation}
\end{center}
We make the assumption that all GIS volume lost makes its way into the oceans.

The contribution to sea level from glaciers and small ice caps (GSIC) is represented by the GSIC sub-model of the Model for Assessment of Greenhouse-gas Induced Climate Change (MAGICC) \cite{wigley2005}. The GSIC sea level contribution ($S_{GSIC}$) is parameterized as
\begin{center}
\begin{equation}
\label{eq:gsic}
\frac{dS_{GSIC}}{dt}(t) = \beta_0 (T_g(t) -T_{eq,GSIC}) \left( 1- \frac{S_{GSIC}(t)}{V_{0,GSIC}} \right)^n.
\end{equation}
\end{center}
In Equation~\ref{eq:gsic}, the uncertain model parameters are: $\beta_0$, the GSIC mass balance sensitivity to global temperature anomalies (m ~\textdegree C$^{-1}$ y$^{-1}$); $V_{0,GSIC}$, the initial GSIC volume susceptible to melt (m SLE); and $n$, the area-to-volume scaling parameter (unitless). These parameters are estimated as in Wong et al. \cite{Wong2017}. $T_{eq,GSIC}$ is taken equal to -0.15~\textdegree C \cite{wigley2005}.

Our parameterization for sea-level rise due to thermal expansion was originally formulated for global sea level by Grinsted et al. \cite{Grinsted2010} and adapted for thermal expansion by Mengel et al. \cite{mengel2016}. First, an equilibrium thermal expansion is calculated ($S_{eq,TE}$), given the anomaly in global mean temperature:
\begin{center}
\begin{equation}
\label{eq:seq_te}
S_{eq,TE}(t) = a_{TE} T_g(t) + b_{TE}.
\end{equation}
\end{center}
$a_{TE}$, the sensitivity of this equilibrium thermal expansion to temperature changes (m~\textdegree C$^{-1}$), and $b_{TE}$, the equilibrium thermal expansion for zero temperature anomaly (m SLE), are estimated as uncertain model parameters \cite{Wong2017}. The thermal expansion contribution to global mean sea level is modeled as an exponential relaxation towards $S_{eq,TE}$:
\begin{center}
\begin{equation}
\label{eq:s_te}
\frac{dS_{TE}}{dt}(t) = \frac{1}{\tau_{TE}} (S_{eq,TE}(t) - S_{TE}(t)),
\end{equation}
\end{center}
where $\tau_{TE}$ is the e-folding time-scale of the thermal expansion response, and the quantity 1/$\tau_{TE}$ is estimated as a model parameter \cite{Wong2017}.

The Antarctic ice sheet is represented by the Danish Center for Earth System Science Antarctic Ice Sheet model, or DAIS~\cite{shaffer2014}. The main equation of state for Antarctic ice sheet volume ($V_{AIS}$, m$^3$) is
\begin{center}
\begin{equation}
\label{eq:ais}
\frac{dV_{AIS}}{dt}(t) = B_{tot}(T,R) + F(S,R),
\end{equation}
\end{center}
where $B_{tot}$ (m$^3$ y$^{-1}$) represents the total rate of accumulation of Antarctic ice sheet mass and $F$ (m$^3$ y$^{-1}$) is the ice volume flux across the grounding line. $T$ is the Antarctic surface temperature reduced to sea level (~\textdegree C), $S$ is sea level (m), and $R$ is the Antarctic ice sheet radius (m). The interested reader is directed to Shaffer \cite{shaffer2014} and Ruckert et al. \cite{Ruckert2017} for more information about the DAIS model.

\subsection{Antarctic Ice Sheet Fast Dynamics Parameterization}
\paragraph{}
The original DAIS model includes a parameterization for dynamic ice loss over the grounding line as it retreats due to subsurface ocean warming ($F$ in Equation~\ref{eq:ais} above; ~\cite{shaffer2014}). This ice flux depends on the Antarctic ice sheet geometry, the water depth, and water temperature. This misses the critical link between rising global temperatures and the sudden, fast ceasing of buttressing ice shelves due to processes such as hydrofracturing and ice cliff failure~\cite{Ruckert2017}, which may substantially speed up the dynamic outflow~\cite{pollard2015}. We form an explicit link between global surface temperatures and these fast dynamical Antarctic contributions to sea-level rise. We parameterize the AIS ``fast dynamics'' disintegration following Diaz and Keller~\cite{diaz2016wais} (their Appendix A):
\begin{center}
\begin{equation}
\label{eq:1} 
\frac{dV}{dt} = \Big\{ \begin{matrix} 
-\lambda, & T > T_{crit} \\ 
0, & T \leq T_{crit}  
\end{matrix}
\end{equation}
\end{center}
where $T_{crit}$ (~\textdegree C) and $\lambda$ (mm yr\textsuperscript{-1}) are uncertain model parameters representing the threshold annual mean temperature at which fast dynamics disintegration occurs and the rate of this disintegration, respectively.  $T$ is the annual mean Antarctic surface temperature, reduced to sea level. Equation~\ref{eq:1} is incorporated as an additional mass balance term into the DAIS model. The parameterization of Equation~\ref{eq:1} represents the bulk contributions from Antarctic ice cliff instability and hydrofracturing to rising sea level. This neglects the causal relationship between (for example) rising temperatures, warming oceans, and sub-ice shelf ocean circulation and these fast processes. Thus, $T_{crit}$ may be thought of as the global warming that coincides with the triggering of the fast ice sheet disintegration processes, but we note the limitation of our formulation to capture only the coincidental relationship, but not the causal. In light of this caveat, for brevity we refer to $T_{crit}$ as the ``trigger temperature'' for the fast dynamics emulator. The DAIS model (without fast dynamics) is described in detail by Shaffer~\cite{shaffer2014}, and the skill of the calibrated DAIS model is described by Ruckert et al.~\cite{Ruckert2017}. 

\paragraph{}
The process approximated by Equation~\ref{eq:1} stops if either the temperature $T$ falls back below $T_{crit}$ or the Antarctic ice sheet volume decreases below 18 million km$^3$. This lower limit is based on the ``extreme interglacial forcing'' scenario of Pollard and DeConto~\cite{pollard2009} and scaling by assumed modern-day Antarctic ice volume (24.78 million km$^3$) and sea level equivalent (57 meters)~\cite{shaffer2014}. Thus, we assume that all ice volume in excess of 18 million km$^3$ is susceptible to fast dynamical collapse.

\paragraph{}
The two-parameter model of Equation~\ref{eq:1} is sufficiently simple that it may be constrained by a paleo record (described below), where the fast dynamics may have occurred either zero or one time. A more complex model would pose considerable computational challenges to constrain observationally. The simple formulation suffices to capture the bulk dynamics of the AIS rapid disintegration, but has limitations. For example, more detailed modeling could consider a probabilistic treatment of the different time-scales, rates and relative contributions from different Antarctic basins susceptible to fast dynamical disintegration~\cite{Ritz2015}. This limitation of our model could lead to unrealistically large contributions to sea-level rise from, say, the West Antarctic ice sheet, which recent work has shown may contribute up to several meters~\cite{pollard2015}. Additionally, our parameterization is calibrated (see Model Calibration, below) to match paleoclimate data assuming an immediate ice sheet response to temperature forcing, which may not be the case. Uncertainty in ice sheet response time-scales likely will induce a wider range of uncertainty in our calibrated estimates of the trigger temperature. Other possible formulations for the Antarctic fast dynamics disintegration might include explicit dependence on the grounding line, for example, as its retreat is driven by rising ocean temperatures. This is a useful avenue for future study, but key strengths of the present approach include (i) it permits estimation of the trigger temperature, $T_{crit}$, and (ii) its simplicity hopefully leads to a transparent analysis of impacts.

\subsection{Model Calibration}
\paragraph{}
The essence of the model calibration approach used here is to update the prior probability distribution of model physical and statistical parameters by quantifying the goodness-of-fit between model hindcasts and observational data. The likelihood function quantifies this match, accounting for uncertainty in each. The posterior distribution of model parameters is given by Bayes' theorem as proportional to the product of the parameters' prior distribution and the likelihood function, evaluated for the model hindcast simulated at the parameter values in question. The model calibration method proceeds by constructing a Markov chain of model parameter estimates, which theoretically converges to samples from the parameter posterior distribution. These samples may be viewed as parameters which yield model simulations that are consistent with observations, given the uncertainty inherent in each. 

\paragraph{}
The substantial parametric uncertainty surrounding the Antarctic fast dynamics contribution to sea-level rise is characterized using two sets of prior distributions for the fast dynamics parameters ($\lambda$ and $T_{crit}$) and running this model calibration algorithm using both sets of fast dynamics priors. We use truncated uniform and gamma distributions for the two sets of priors. For the truncated uniform priors, $\lambda$ ranges from 5 mm yr$^{-1}$ to 15 mm yr$^{-1}$, centered at a recent estimate~\cite{deconto2016}; the range for $T_{crit}$ (in Antarctic surface temperature reduced to sea level) is from -20~\textdegree C to -10~\textdegree C. The parameters for the gamma priors are chosen to keep the mean at the center of the uniform priors, and place the 5\% quantile for $\lambda$ at 5 mm yr$^{-1}$ and for $T_{crit}$ at -10~\textdegree C. The prior distributions for all other model parameters are the same between the two experiments (see \nameref{Online_Resource_2}).

\paragraph{}
We construct paleoclimatic calibration windows for the Last Interglacial (118,000 years before current era (BCE))~\cite{deconto2016}, Last Glacial Maximum, mid-Holocene, and instrumental period~\cite{shaffer2014,Ruckert2017,Wong2017}. These windows are combined with AIS mass loss trends from the International Panel on Climate Change (IPCC) AR5~\cite{ipcc2013wg1} to constrain the Antarctic ice sheet simulation. The Last Interglacial window uses a truncated normal likelihood function between 3.6 and 7.4 m sea-level equivalent (SLE)~\cite{deconto2016}, with mean 5.5 m and standard deviation 0.95 m. A Heaviside likelihood function is also used for the total sea-level rise due to the Antarctic ice sheet, as well as the thermal expansion, to exclude simulations that yield individual components of sea-level rise which exceed the total sea-level rise data. The other paleoclimatic calibration periods use Gaussian likelihood functions. The date, mean, and standard deviation of these are (respectively): 18,000 years BCE, -11.35 m SLE, 2.23 m SLE; 4,000 years BCE, -2.63 m SLE, 0.68 m SLE; and 2002 CE, 0.00197 m SLE, 0.00046 m SLE~\cite{Ruckert2017}. The paleoclimatic calibration runs span 240,000 years before current era to present.

\paragraph{}
Other observational data used to constrain the model parameters include global mean surface temperature~\cite{morice2012}, ocean heat uptake~\cite{gouretski2007}, glaciers and small ice caps~\cite{dyurgerov2005}, Greenland ice sheet~\cite{sasgen2012}, thermal expansion trends from the IPCC AR5~\cite{ipcc2013wg1}, and global mean sea level~\cite{church2011}. We implement a simple, first-order autoregressive (``AR1'') error model for the model-data residuals for the surface temperature, ocean heat uptake, glaciers and small ice caps, and Greenland ice sheet. These error models include homoscedastic error ($\sigma$) and autocorrelation ($\rho$) statistical parameters for each component. Median timescales $T$ (years) on which the temperature, ocean heat, glacier and ice cap, and Greenland ice sheet residuals become uncorrelated (lag-$T$ autocorrelation coefficient $< 0.05$) are 5, 9, 6, and 8 years, respectively. Of course, longer timescale (e.g., multi-decadal) modes in these time series are present (particularly in the ocean heat), but are not of interest to the present study. Additionally, we include heteroscedastic error estimates for the temperature, ocean heat uptake, and glaciers and small ice caps data, adding the homoscedastic and heteroscedastic error components in quadrature. 

\paragraph{}
The non-Antarctic ice sheet model components (modern calibration) and Antarctic ice sheet model (paleoclimatic calibration) are calibrated separately using a robust adaptive Metropolis Markov chain Monte Carlo (MCMC) algorithm~\cite{vihola2012}. This algorithm adapts the covariance matrix of the multivariate Gaussian distribution used to propose new parameter iterates, centered at the current parameter estimates. This method takes into account the correlation structure of previous parameter iterates. Four parallel Markov chains of 1,000,000 iterations each for the modern calibration and of 500,000 iterations each for the paleoclimatic calibration are generated. Gelman and Rubin diagnostics are evaluated to assess convergence~\cite{gelman1992}. The first 500,000 iterations of each of the modern calibration Markov chains and the first 300,000 iterations of the paleoclimatic calibration Markov chains are discarded for burn-in. This yields posterior samples of 2,000,000 and 800,000 parameter sets for the modern and paleoclimatic calibration parameters, respectively.

\paragraph{}
From each of the two disjoint resulting posterior samples, 30,000 random samples of model parameters are drawn and combined into sets to run the full model (AIS and non-AIS). The full model was run from 1850 to present at these parameter samples and calibrated to total global mean sea-level rise data~\cite{church2011} using rejection sampling. Contributions from land water storage were subtracted out in a preliminary step, using IPCC AR5 trends and adding the uncertainties in sea level and land water storage in quadrature~\cite{ipcc2013wg1}. This step assumes closure of the global sea level budget, which while not always strictly true throughout the instrumental period, is a reasonable assumption from 1900 onward~\cite{ipcc2013wg1}. The enveloping distribution for rejection sampling is the joint Gaussian likelihood function for the sea level data (corrected for land water storage), evaluated at the observed sea level time series itself (since the likelihood function for any model simulation cannot exceed this value). Model simulations are accepted with probability equal to the ratio of the likelihood function evaluated at the selected model simulation to the maximal value of the likelihood function. This sea level calibration results in ensembles for analysis of 2,867 and 2,850  members for the uniform and gamma prior experiments, respectively. \nameref{Online_Resource_1} provides the calibrated marginal distributions for all model parameters, for both sets of priors. Ensembles of projections for each of Representative Concentration Pathways (RCP) 2.6, 4.5, and 8.5 are generated using the same calibrated parameters. This yields six projected sea-level rise scenarios: three forcing scenarios times two fast dynamics prior assumptions. We only present the results for the gamma priors here; both sets yielded similar projections of sea-level rise and AIS fast dynamical disintegration (see \nameref{Online_Resource_1}).

\subsection{Local Coastal Defense}
\paragraph{}
The ensembles of global mean sea-level rise are converted to local sea-level rise for New Orleans, Louisiana, using previously published regional scaling factors for each component of global sea-level rise~\cite{slangen2014}. We assume local sea level fingerprints of 0.89 for glaciers and ice caps, 1.1 for the Antarctic ice sheet, and 0.81 for the Greenland ice sheet. We make the assumption that thermal expansion of the oceans affects local sea levels uniformly and use a fingerprint of 1.0 for this contribution. In light of the lack of specific information regarding local contributions of land water storage, we use a fingerprint of 1.0 for land water storage as well. Preliminary experiments suggest that our results are not sensitive to the specific values for the fingerprints for land water storage and thermal expansion. Given these local sea-level rise projections for this century, we perform an economic optimization for flood safety levels of New Orleans, Louisiana. The essence of this approach is to balance the net present value of the costs associated with both (i) investing in greater levels of flood protection through levee heightening (at the starting year) and (ii) the losses from flood damages due to inadequate levels of protection, given the sea-level rise realizations and associated flood probabilities for a given levee height~\cite{dantzig1956,jonkman2009b}. We consider cases with and without accounting for the fast dynamics contribution to sea-level rise to assess the impacts of this mechanism on coastal defense strategies.  Flooding occurs only through water levels overtopping the levee; structural failure is not considered here nor in the original analyses~\cite{dantzig1956,jonkman2009b}, but is an interesting avenue for further study. 

\paragraph{}
The cost-benefit analysis assumes the current year is 2015 and considers a time horizon of 2100 (85 years). Levee heightenings (at the starting year) between 0 and 10 meters are considered, in increments of 5 centimeters. The average annual flood probability is calculated for each proposed levee heightening from the simulated local sea-level rise, the land subsidence rate~\cite{dixon2006} , and flood frequency parameters~\cite{dantzig1956} following the method outlined in Van Dantzig~\cite{dantzig1956}. Local subsidence at New Orleans is attributable to a range of factors, including (for example) the extraction of groundwater, oil and gas, sediment compaction, faulting, and glacial isostatic adjustment \cite{Jones2016}. The rate of land subsidence follows a log-normal distribution (to prevent unreasonable negative values) with mean 5.6 mm y$^{-1}$ and standard deviation 0.4 mm y$^{-1}$, based on high-resolution satellite measurements~\cite{dixon2006}. The flood probability ($p_f$) is distributed exponentially with respect to sea level above the levee height, with the rate constant $\alpha$ considered as an uncertain parameter (Table \ref{tab1}):
\begin{center}
\begin{equation}
\label{eq:pf} 
p_f = p_0 e^{-\alpha (\Delta h - \Delta S)}\ .
\end{equation}
\end{center}
In Equation~\ref{eq:pf}, $\Delta h$ is the proposed levee heightening (m), $p_0$ is the flood probability with zero additional heightening (Table \ref{tab1}), and $\Delta S$ is the local mean sea-level rise (m).

\begin{table}[hbt]
\newlength{\digitwidth} \settowidth{\digitwidth}{\rm 0}
\catcode`?=\active \def?{\kern\digitwidth}
\caption{Parameters for flood protection cost-benefit analysis and their sampling distributions.}
\label{tab1}
\begin{tabular*}{\textwidth}{@{}l@{\extracolsep{\fill}}lll}
\hline
Parameter		& \ \ Description & Distribution \\
\hline
$p_0$			& \shortstack[l]{Initial flood frequency (yr$^{-1}$)\\with zero heightening} & $\log{N}(\log{\mu}=\log(0.0038), \log{\sigma}=0.25)$ \\
$\alpha$		& \shortstack[l]{Exponential flood\\frequency constant (m$^{-1}$)} & $N(\mu=2.6$, $\sigma=0.1)$ \\
$V$					& \shortstack[l]{Value of goods protected by\\levee polder (billion US\$)} & $U(5, 30)$ \\
$\delta$		& Net discount rate (\%) & $U(0.02, 0.06)$ \\
$I_{unc}$		& Investment uncertainty (\%) & $U(0.5, 1)$ \\
$r_{subs}$	& Land subsidence rate (m yr$^{-1}$) & $\log{N}(\log{\mu}=\log(0.0056), \log{\sigma}=0.4)$ \\
\hline
\end{tabular*}
\end{table}

\paragraph{}
These flood probabilities are then combined with the value of goods protected by the levee ring ($V$) and the monetary discount rate $\delta$ (Table \ref{tab1}; \cite{jonkman2009b}) to calculate expected losses (US dollars) for each proposed levee heightening. The expected losses are then
\begin{center}
\begin{equation}
\label{eq:loss} 
L(\Delta h) = p_0 e^{-\alpha (\Delta h - \Delta S)} \frac{V}{(1+\delta)^t} \ ,
\end{equation}
\end{center}
where $t$ is the future time to which the value $V$ is discounted. The expected investments ($I(\Delta h)$) are approximated as a linear function of the proposed heightening, using cost estimates from previous studies~\cite{jonkman2009b}. The total costs are the sum of the expected investments and the expected losses, $C(\Delta h) = L(\Delta h) + I(\Delta h)$, and the economically-efficient levee heightening is the value $\Delta h$ that minimizes $C(\Delta h)$. 

\begin{figure}[!ht]
\begin{center}
\includegraphics[width=70mm]{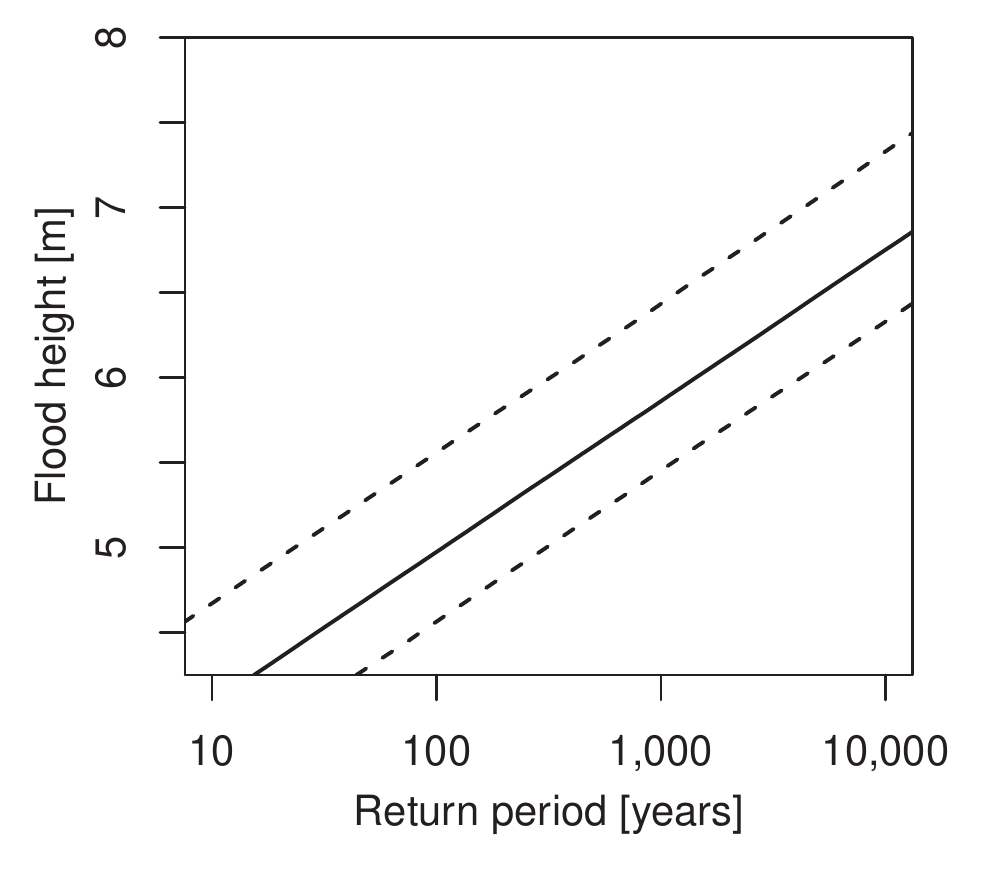}
\caption{{\bf Relationship between return period (reciprocal of the average exceedance probability over the 2015-2100 time period) and flood height for the ensemble median (solid line) and 5-95\% credible range (dashed lines) under RCP8.5~\cite{meinshausen2011rcp}.}}
\label{fig0}
\end{center}
\end{figure}

The ``return period'' corresponds to the frequency of storms with the potential to overtop levees with the corresponding levee height (Figure~\ref{fig0}). For example, a 100-year return period corresponds to a 1/100 average annual flood probability (or the 1:100 level of protection). For a given investment in levee heightening, if fast dynamics are neglected the return period is shorter than the return period if fast dynamics are included. This is because the additional contributions of sea-level rise lead to a realized return period that is shorter than the presumed (or goal) return period.

\paragraph{}
Each sea-level rise ensemble member is assigned a corresponding set of parameters for the flood risk analysis. We present only the ensemble under the RCP8.5 radiative forcing. Uncertainty in the parameters for the cost-benefit analysis is incorporated using a Latin hypercube sample and parameter ranges given by Table \ref{tab1}. The parameter ranges were selected to capture the sensitivity of the coastal defense cost-benefit analysis to these uncertain parameters as in other recent analyses~\cite{jonkman2009b}. We consider only the north-central levee ring in New Orleans (see, for example, ~\cite{jonkman2009b}, their Fig. 1). It is important to recall that this illustration still neglects key processes and uncertainties (for example, storm surges and structural failure besides overtopping), and is not to be used to inform on-the-ground decisions.

\section{Results}
\subsection{Model Hindcast}
\begin{figure}[!ht]
\begin{center}
\includegraphics[width=.85\linewidth]{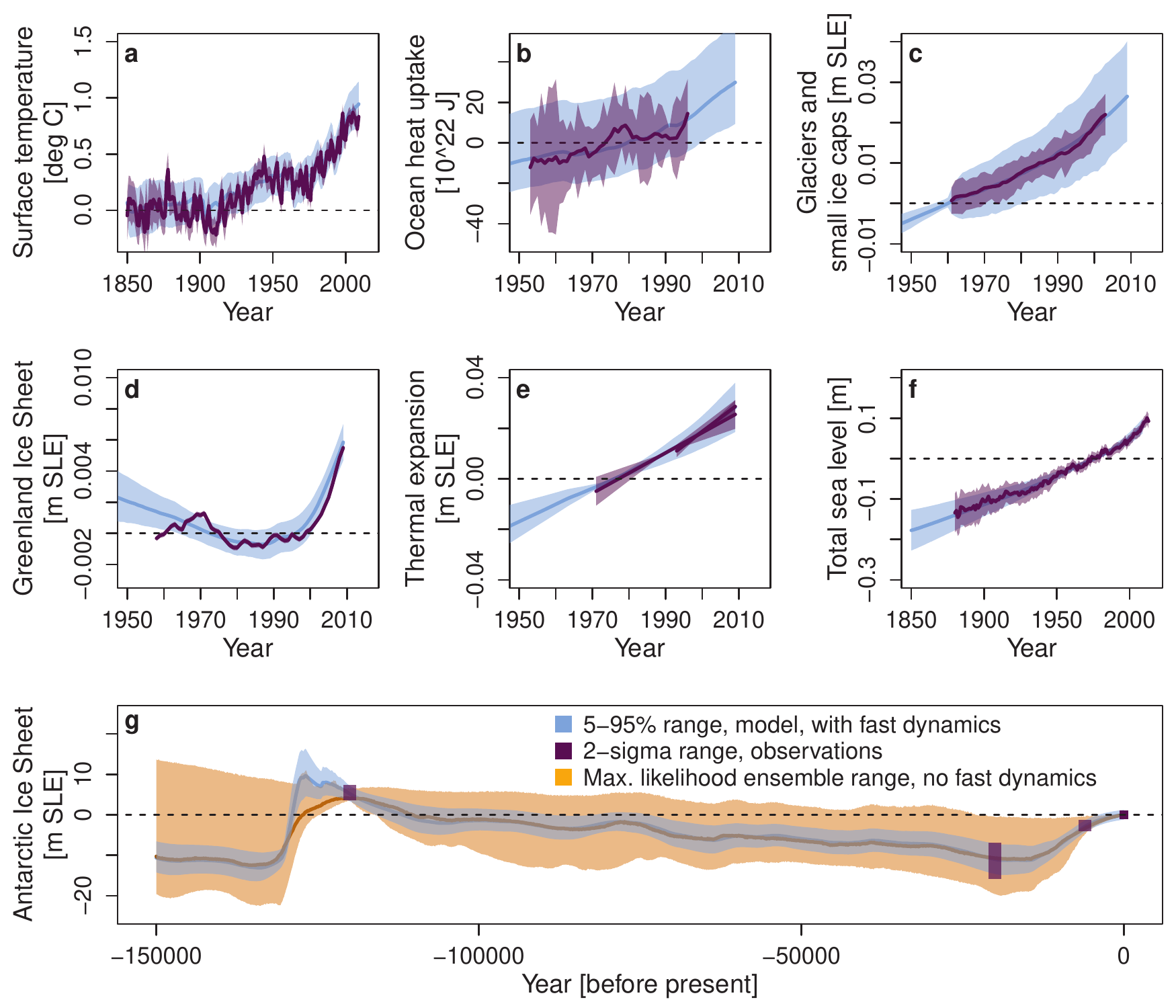}
\caption{{\bf Modeled (blue) and observed (purple) calibration data fields. The thick lines indicate the ensemble medians or the observations. The shaded regions indicate a 5-95\% credible range (model) or a 2$\sigma$ range (observations). (a) surface temperature anomaly (C), (b) ocean heat uptake (10$^{22}$ J), (c) glaciers and small ice caps (m), (d) Greenland Ice Sheet (m), (e) thermal expansion (m), (f) total sea level (m), and (g) Antarctic Ice Sheet (m).}}
\label{fig1}
\end{center}
\end{figure}

\paragraph{}
The hindcast skill of the BRICK platform of models used here, run at fully calibrated parameter sets is demonstrated in Fig. \ref{fig1} (see also \nameref{Online_Resource_1} and 2). The model ensemble after calibration reproduces the central statistics of the data well (darkened lines represent the ensemble median time series) and also reproduces the ranges seen in the observational data (light shaded regions represent the 5-95\% ranges in the model ensemble, and 2$\sigma$ range about the observational data). 

\newpage

\paragraph{}
Fig. \ref{fig1}g also includes an ensemble of Antarctic ice sheet simulations in which the fast dynamics emulator is not enabled. This ensemble is constructed using a Latin hypercube sample of the AIS model parameters. We use a Latin hypercube sampling approach because without the fast dynamics emulator, the Bayesian calibration algorithm fails to converge. The ensemble consists of the 10\% highest realizations of the likelihood function (i.e., the 10\% ``most likely'' model simulations). The weak constraint on these simulations from the paleoclimate observations is attributed to both the inefficient calibration method used for this specific illustration and the lack of the fast dynamics mechanism. It is particularly illuminating that the statistical calibration method fails when key physics (i.e., the fast dynamics) are neglected. The period leading up to the Last Interglacial calibration window (118,000 years BCE) is the only period during which the more tightly constrained ensemble that includes fast dynamics exceeds the ensemble that does not include fast dynamics (Fig. \ref{fig1}g). This demonstrates that elevated global surface temperatures during this period are driving the paleoclimatic Antarctic fast dynamical sea level contributions.

\subsection{Global Warming Triggering Fast Antarctic Disintegration}
\begin{figure}[!ht]
\begin{center}
\includegraphics[width=84mm]{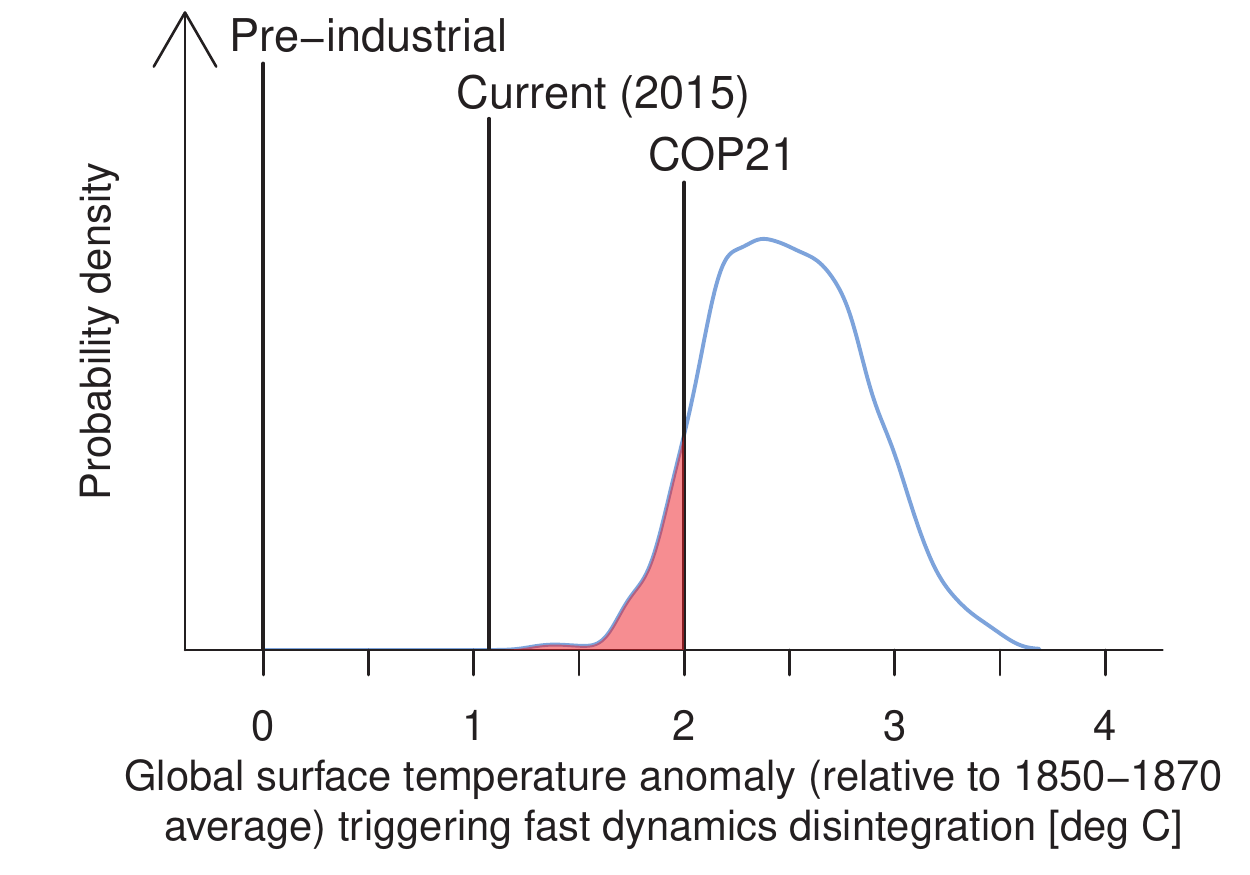}
\caption{{\bf Calibrated distributions of the trigger temperature for Antarctic fast dynamics contribution to sea-level rise, relative to the 1850--1870 global mean surface temperature. The pre-industrial (1850--1870 mean), current (2015), and 2~\textdegree C COP21 resolution~\cite{rhodes2016} temperature are shown as vertical lines.}}
\label{fig2}
\end{center}
\end{figure}

\paragraph{}
We find that the trigger temperature of the AIS disintegration ($T_{crit}$) is reasonably well-constrained by the paleoclimate data (Fig. \ref{fig2}). This conclusion is based on the fact that the period just before the Last Interglacial is the only time during which the fast dynamics mechanism is triggered, so the paleoclimatic record provides the constraint on the distribution of $T_{crit}$ (see also \nameref{Online_Resource_1}). The resulting estimate for $T_{crit}$ is 2.5~\textdegree C (ensemble median; 5-95\% range is 1.9-3.1~\textdegree C). This trigger temperature has been scaled from Antarctic mean surface temperature to global mean surface temperature anomaly (relative to 1850-1870 mean) using paleoclimate reconstructions~\cite{morice2012,shaffer2014}. The relationship between global and Antarctic local temperatures is complex and uncertain. Thus, the uncertainty in the distribution of $T_{crit}$ as a global mean temperature is likely higher, leading to a wider distribution than is found here. In light of this caveat, even when the global temperature remains below the 2~\textdegree C warming target from the recent Paris Agreement~\cite{rhodes2016}, there is still possibility to trigger the AIS dynamics, according to this simple analysis. The total probability below 2~\textdegree C warming is approximately 9\% (Fig. \ref{fig2}, shaded red region). By contrast, the total probability below 1.5~\textdegree C warming is substantially lower, at 0.3\%.

\begin{figure}[!ht]
\begin{center}
\includegraphics[width=84mm]{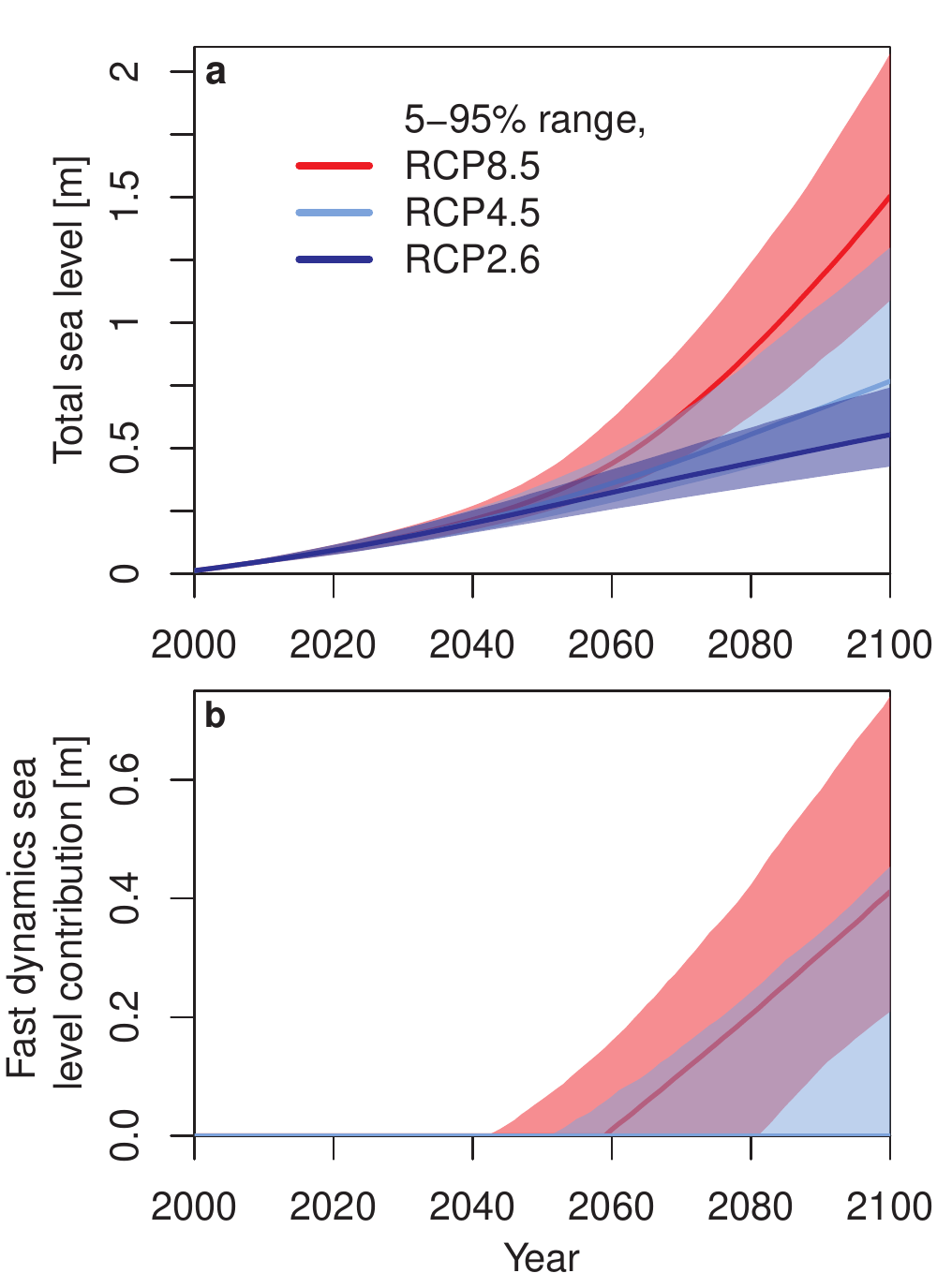}
\caption{{\bf Projections of (a) total sea level and (b) the Antarctic fast dynamics contribution to sea-level rise, relative to the global mean 1986-2005 sea level, under RCP2.6 (blue), RCP4.5 (light blue), and RCP8.5 (red)~\cite{meinshausen2011rcp}.}}
\label{fig3}
\end{center}
\end{figure}

\paragraph{}
Under RCP8.5, we find that the fast dynamics contribution to sea-level rise in 2100 is 41 cm (ensemble median; 5-95\% range is 21-74 cm, Fig. \ref{fig3}b). The median year in which the AIS fast dynamics disintegration initiates is 2060 (5-95\% range is 2043-2082) under RCP8.5. Under RCP4.5, the ensemble median in 2100 does not include any disintegration, but contributions up to 45 cm are possible (95\% quantile). Under RCP2.6, to the 95\% credible level, no disintegration occurs. 

\begin{figure}[!ht]
\begin{center}
\includegraphics[width=84mm]{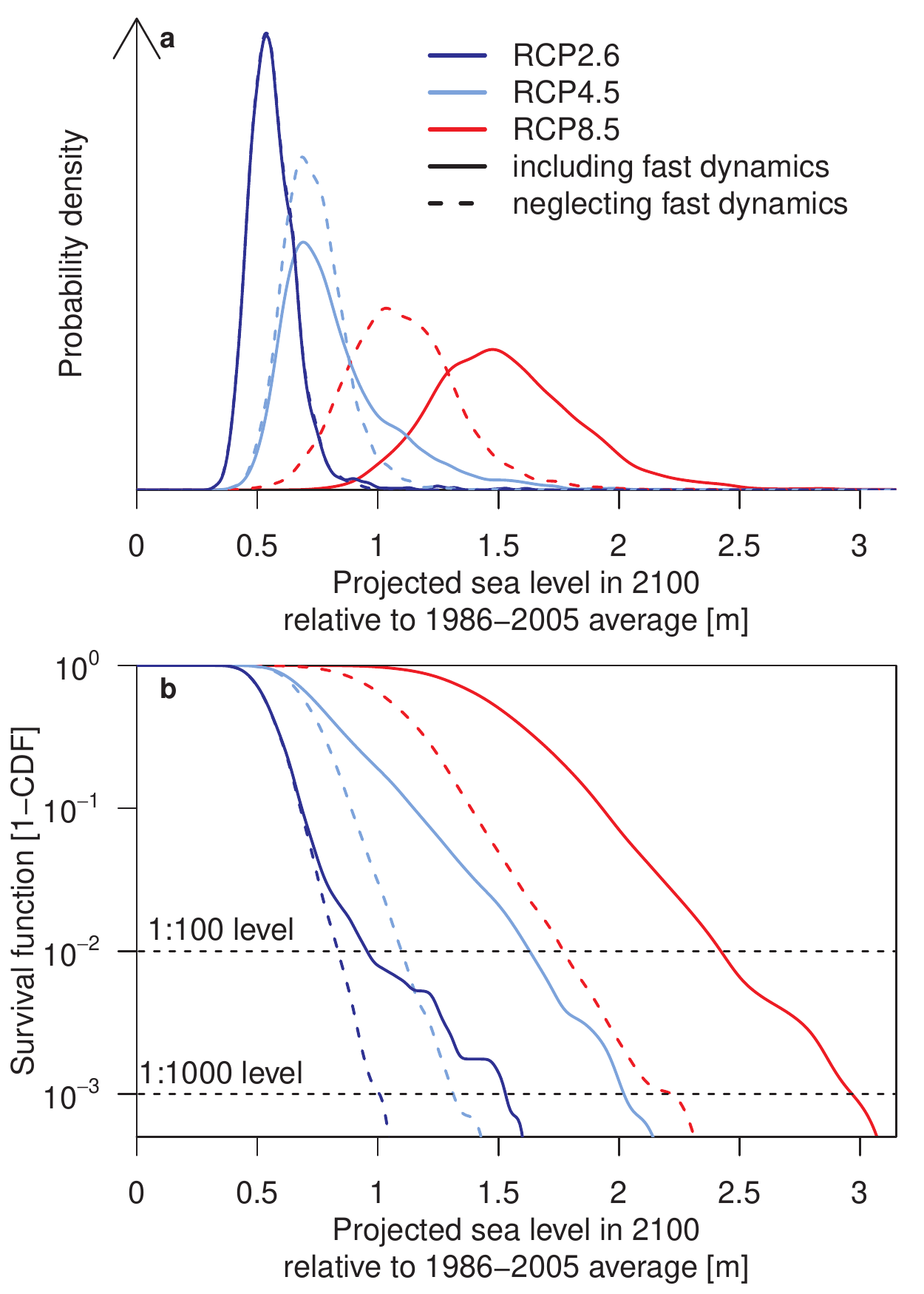}
\caption{{\bf Distribution of projected sea level in 2100 with (solid lines) and without (dashed lines) accounting for the Antarctic fast dynamics contribution, under RCP2.6 (blue), RCP4.5 (light blue), and RCP8.5 (red)~\cite{meinshausen2011rcp}. (a) Probability density functions and (b) survival functions (which give the total probability in the right tail of a distribution).}}
\label{fig4}
\end{center}
\end{figure}

\paragraph{}
The fact that a medium forcing (RCP4.5) does not trigger disintegration in the ensemble median (Fig. \ref{fig4}b) is not an indicator of safety. In fact, the probabilistic projections (Fig. \ref{fig4}) show that ignoring the fast dynamics sea-level rise leads to neglecting relevant low-probability but high-impact events. Under RCP4.5, the 2100 sea level displays a substantial tail above one meter, whereas neglecting fast dynamics completely misses this potentially large sea-level rise (Fig. \ref{fig4}a). Even under RCP2.6, the fast dynamics sea-level rise is noticeable beyond the 1:100 level (Fig. \ref{fig4}b). Under RCP4.5 and 8.5, it can be seen that at typically applied reliabilities (e.g., 1:100), the fast dynamics cannot be ignored. 

Medians and 5-95\% ranges for total sea-level rise in 2100 are 55 cm (43-74 cm, RCP2.6), 77 cm (56-130 cm, RCP4.5) and 150 cm (109-207 cm, RCP8.5). We find the Antarctic (including fast dynamics) contribution to these projections to be 9 cm (2-16 cm, RCP2.6), 11 cm (3-50 cm, RCP4.5) and 44 cm (24-80 cm, RCP8.5). These projections are lower than those of DeConto and Pollard \cite{deconto2016}, whose (for example) highest ensemble estimate of Antarctic contribution to sea-level rise by 2100 is 114$\pm$36 cm (RCP8.5, relative to sea level in 2000). This result is not surprising given our simple model coupled to a detailed calibration approach, versus their detailed model/simple calibration approach. We address this further in the Discussion.

\subsection{Implications for Coastal Defense}
\begin{figure}[!ht]
\begin{center}
\includegraphics[width=\linewidth]{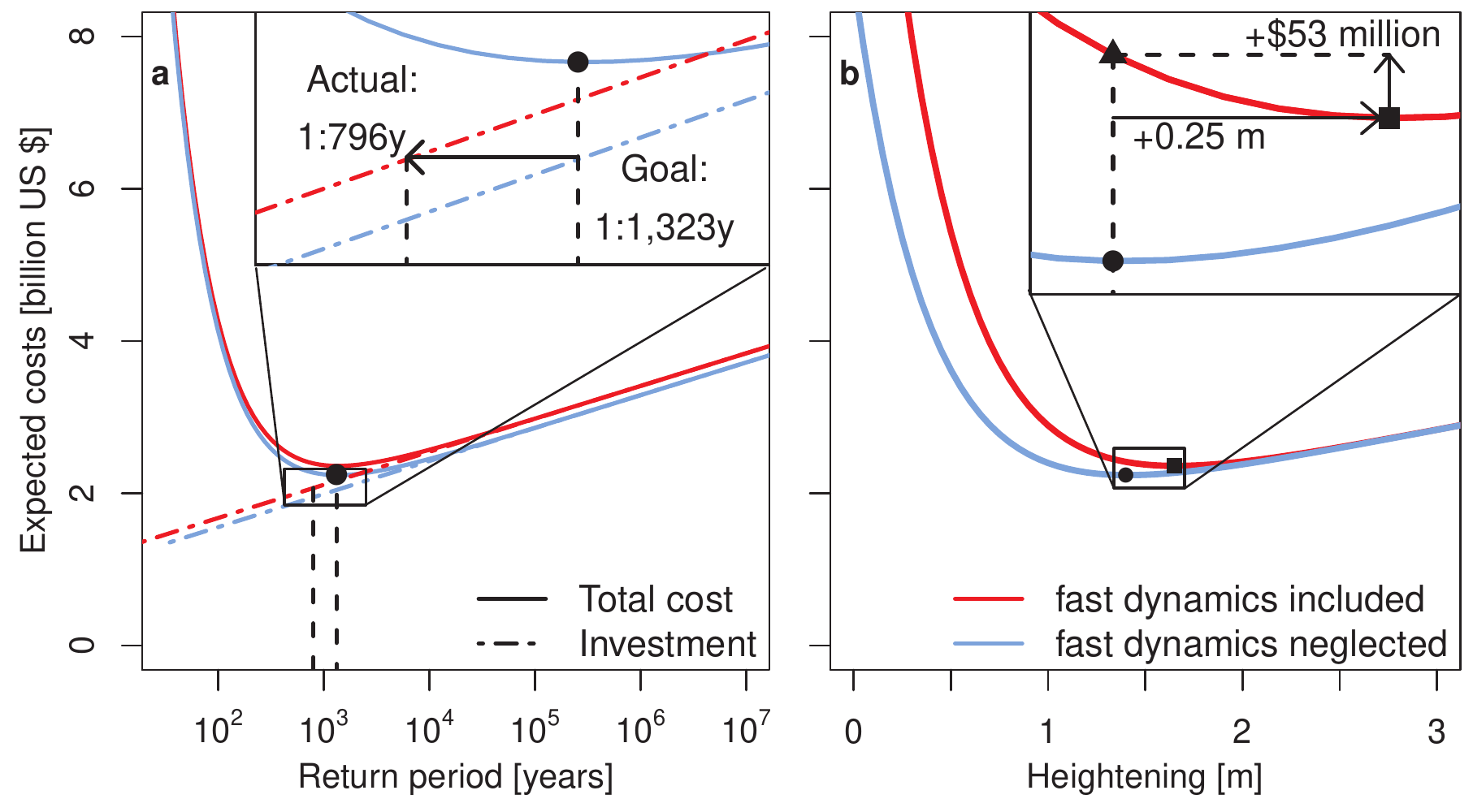}
\caption{{\bf Illustrative cost-benefit analysis of (a) the optimal (i.e., economically efficient) return period and (b) the levee heightening for the north-central levee ring in New Orleans, Louisiana. The symbols denote the optimal strategy assuming no knowledge of the fast dynamics sea-level rise (filled circle); accounting for the fast dynamics (filled square); and the if the optimal strategy that neglects fast dynamics encounters sea-level rise that includes fast dynamics contributions (filled triangle).}}
\label{fig5}
\end{center}
\end{figure}

\paragraph{}
Adopting sea-level projections that neglect the Antarctic fast dynamics yield an economically efficient return period of about 1,300 years (Fig. \ref{fig5}a, under RCP8.5). This corresponds to a levee heightening of 1.4 meters (Fig. \ref{fig5}b). Confronting such a levee with the arguably more realistic sea-level rise projections that include the fast dynamics, the level of protection achieved drops to a return period of about 800 years (Fig. \ref{fig5}a, inset). This increase in flood risk (the inverse of return period) is due to the additional hazard posed by the fast dynamical Antarctic contributions to sea-level rise. With consideration of fast AIS dynamics, the economically efficient levee heightening is 1.65 meters, with a return period of roughly 1,300 years. This is lower than the 5,000-year economically-efficient return period reported for this levee ring by Jonkman et al. \cite{jonkman2009b}. The Louisiana Coastal Protection and Restoration Authority (CPRA) has protection targets of 100 year return period for standard construction projects and 500 years for critical infrastructure such as hospitals~\cite{CoastalProtectionandRestorationAuthorityofLouisiana2017}. Our results suggest that these protection standards may not be economically efficient, especially considering (1) that our analysis neglects the effects of storm surges and (2) that the 50-year planning period considered by the CPRA overlaps considerably with estimates of the timing of Antarctic fast disintegration, both presented here and elsewhere~\cite{Ritz2015,deconto2016}.

\paragraph{}
We calculate the total expected costs of the heightening strategy neglecting fast dynamics (solid circle, Fig. \ref{fig5}) when this strategy is confronted by sea-level rise that includes fast dynamics (solid triangle, Fig. \ref{fig5}b). Accounting for the fast dynamics reduces the total expected costs in this simple analysis by \$53 million (solid square versus triangle, Fig. \ref{fig5}b). Under RCP4.5, the ensemble mean economically efficient heightening with the fast dynamics is only 3 centimeters taller than the efficient heightening without the fast dynamics, with a mean reduction in expected costs of \$2.1 million. Under RCP2.6, the two strategies typically do not differ because the fast dynamics are not triggered in most simulations; the ensemble mean additional heightening in consideration of the fast dynamics is 0.2 centimeters.

\section{Discussion and Caveats}
\paragraph{}
Our analysis should be interpreted as an illustrative example using a simple model. This simple approach results in a hopefully transparent analysis, but also gives rise to important caveats. For example, we analyze just one levee ring, use a simple economic model, and neglect many uncertainties and processes (e.g., structural failure or changes in future storm surges~\cite{moritz2015}). Additionally, the probabilistic projections and analysis presented here focus on a relatively short time horizon compared to the committed sea-level response. For example, even under RCP4.5, near-complete disintegration of the WAIS is possible by 2500~\cite{deconto2016}, so extending the projections to 2500 is an path for future study (see \nameref{Online_Resource_1} for projections to 2200). 

\paragraph{}
The total value of assets assumed to be protected by the levee system is \$5 to 30 billion (Table \ref{tab1}). Thus, the \$53 million reduction in expected damages by including the fast dynamics contributions to sea level in the flood protection is less than about 1\% of the total value of assets. Note that this seemingly low figure must be carefully balanced against the numerous considerations it for which it does not account. These include potential future losses due to saltwater intrusion, the cultural significance of damaged assets or areas~\cite{bessette2017}, as well as the loss of life and associated future economic losses. Additionally, the model for flood risk assessment employed here assumes that the levee system is heightened instantaneously upon evaluation (see Section 2.4). Future studies should consider a multi-stage adaptive design approach, wherein the levee system is reevaluated at specific intervals but cannot be heightened by more than a prescribed amount each year.

\paragraph{}
Future work could also expand on the simple parameterization for the Antarctic fast dynamical processes through more complex model structure. For example, our parameterization cannot resolve the individual contributions to the disintegration rate; different Antarctic basins could respond at different temperatures and with different rates. By capturing only an ice sheet average disintegration rate and onset time-scale, our simple model likely overestimates the year in which disintegration may begin and underestimate the disintegration rate relative to more complex models (Ritz et al. 2015; DeConto and Pollard 2016). Indeed, the results of Section 3.2 lend credibility to this hypothesis relative to DeConto and Pollard (2016). Our estimates for the timing of the onset of fast disintegration (2043 is the 5\% quantile under RCP8.5) are quite compatible with the probabilistic timing estimates of Ritz et al. (2015), who find a roughly 5\% probability of exceeding 5 cm of sea level contribution from fast disintegration by 2040 (c.f., their Fig. 2). Our parameterization assumes an immediate ice sheet response to the trigger temperature, which may not be the case in reality. An additional time lag parameter could be incorporated into the parameterization and model calibration framework, although additional data should be included. Potential future data for assimilation may include paleoclimate data from the Pliocene~\cite{deconto2016} as well as expert assessment regarding future Antarctic ice sheet mass loss~\cite{Oppenheimer2016a,bakker2016b}. These approaches hold promise for refining the estimates of the trigger temperature (Fig. \ref{fig2}) as new information becomes available. Fast dynamical disintegration may also be a threshold event, so the possibility of stopping the disintegration by cooling $T$ back below $T_{crit}$ may not be physically realistic. The caveats point to important research needs and also illustrate why the results should not be used to directly inform on-the-ground decisions. 

\paragraph{}
As compared to other probabilistic projections of sea-level rise this century, our estimates are substantially higher but not out of agreement (within the 5-95\% range) of previous work. Under RCP8.5, we estimate sea-level rise of 109 to 207 cm by 2100, as compared to 52 to 131 cm~\cite{kopp2016}, 57 to 131 cm~\cite{mengel2016}, and 37 to 118 cm~\cite{Jackson2016}. This is perhaps not surprising, as these previous projections do not include the fast AIS dynamics in their probabilistic frameworks. The 95\% quantile for sea-level rise by 2100 presented here of 207 cm is roughly consistent with the 95\% quantile of 180 cm reported by Jevrejeva et al.~\cite{Jevrejeva2014}, but notably higher than the 95\% quantile of 121 cm found by Kopp et al. \cite{Kopp2014}. Both of these latter studies combined process-based modeling with expert assessment \cite{bamber2013b} to account for the potential Antarctic fast dynamical sea level contributions. It is likely that improved agreement with Jevrejeva et al. \cite{Jevrejeva2014} stems from their broader accounting of uncertainty in the expert assessment, as compared to Kopp et al. \cite{Kopp2014}.

\section{Conclusions}
\paragraph{}
Given these caveats, we provide calibrated probabilistic sea-level projections, accounting for the AIS fast dynamics using a simple parameterization. Our projections are quite capable of exceeding previous estimates of upper limits on sea-level rise in this century~\cite{pfeffer2008}. The projected time horizon of 2043-2082 (5-95\% range under RCP8.5) for fast dynamics disintegration is in agreement with a recent study which predicts about 2050~\cite{deconto2016}. Our approach differs from theirs in ensemble size and model complexity, yet the resulting time horizons of AIS disintegration are quite similar, which lends credibility to both studies. Our results offer a potential marker for triggering AIS fast disintegration in the form of the calibrated distributions of trigger temperature, $T_{crit}$ (Fig. \ref{fig2}). The 2~\textdegree C increase in global mean surface temperature designated in the Paris Agreement~\cite{rhodes2016} is within the 5-95\% ensemble range of $T_{crit}$ (1.9-3.1~\textdegree C). This indicates that temperature increases within the 2~\textdegree C limit may still lead to Antarctic fast dynamical disintegration. Further, these results demonstrate how lowering emissions can be an avenue to drastically reduce coastal flooding risks.

\paragraph{Acknowledgements} 
We thank Dave Pollard for lending his insight into reasonable prior ranges and central values for the Antarctic ice sheet fast dynamics parameters, and for comments on the initial version of the manuscript. We thank Kelsey Ruckert and Yawen Guan for their assistance in providing and interpreting codes relevant to the original DAIS model and its calibration. We also thank Kelsey Ruckert for manuscript formatting assistance. We thank Aimée Slangen for supplying and assistance interpreting the regional sea level fingerprinting data. We gratefully acknowledge Richard Alley, Murali Haran, Chris and Bella Forest, Rob Nicholas, Patrick Reed, Michael Oppenheimer, Tad Pfeffer, Rob Lempert, David Johnson, Roger Cooke, and Dale Jennings for invaluable inputs. This work was partially supported by the National Science Foundation through the Network for Sustainable Climate Risk Management (SCRiM) under NSF cooperative agreement GEO-1240507 as well as the Penn State Center for Climate Risk Management.  Any conclusions or recommendations expressed in this material are those of the authors and do not necessarily reflect the views of the funding agencies. Any errors and opinions are, of course, those of the authors.

\paragraph{Author contributions}
T.W. and K.K. initiated the study. T.W., A.B., and K.K. designed the research. T.W. and A.B. produced the model simulations. T.W. designed the initial figures and wrote the first draft. All contributed to the final text.

\paragraph{Code availability}
The BRICK model (with the added fast dynamics module) and analysis codes are freely available from \url{https://github.com/scrim-network/BRICK/tree/fastdy}. Large parameter files and model results files are available from \url{https://download.scrim.psu.edu/Wong_etal_BRICK}. Code examples and a routine for fingerprinting sea-level rise projections to other locations (aside from New Orleans, as presented in this manuscript) is provided at \url{https://github.com/scrim-network/BRICK}. The physical models are coded in Fortran 90 and called from driver scripts coded in the R Programming Language. The analysis was performed using RStudio (Version 0.99.903). 

\section*{Electronic supplementary material captions}

\paragraph{Online Resource 1}
\label{Online_Resource_1}
Supplementary figures are provided, with captions.

\paragraph{Online Resource 2}
\label{Online_Resource_2}
Page 1 table contains: parameter names (column 1); model of origin (column 2); ensemble median, 5\% quantile, and 95\% quantile (columns 3, 4, and 5), from the experiment using the gamma priors for the fast dynamics; prior range lower and upper bounds, when assigned uniform priors (columns 6 and 7); and units (column 8). Page 2 table contains: parameter names (column 1); model of origin (column 2); and brief description (column 3).

\bibliography{library}

\bibliographystyle{abbrv}

\end{document}